\documentclass[a4paper,11pt]{article}

\usepackage{tikz}
\usepackage{listings, enumitem, isotope, wrapfig, float}
\usepackage{makecell}
\usepackage{siunitx}
\usepackage{jinstpub} 

\usepackage{courier}
\newcommand{\red}[1]{{#1}}
\newcommand{\e}[2]{$#1 \cdot 10^{#2}$}

\newcommand*{\doi}[1]{\href{https://doi.org/#1}{doi:\,#1}}
\newcommand*{\arxiv}[1]{\href{http://arxiv.org/abs/#1}{arXiv:\,#1}}

\title{\boldmath Software-based data acquisition and processing
for neutron detectors at European Spallation Source --- early
experience from four detector designs.}

\author[a,1]{M. J. Christensen,\note{Corresponding author.}}
\author[a]{M. Shetty,}
\author[a]{J. Nilsson,}
\author[a]{A. Mukai,}
\author[b,c]{R. Al Jebali,}
\author[b]{A. Khaplanov,}
\author[f]{M. Lupberger,}
\author[b,g]{F. Messi,}
\author[b,f]{D. Pfeiffer,}
\author[b]{F. Piscitelli,}
\author[d]{T. Blum,}
\author[d]{C. S\o gaard,}
\author[d]{S. Skelboe,}
\author[b,e]{R. Hall-Wilton,}
\author[b]{T. Richter}

\affiliation[a]{European Spallation Source, Data Management and Software
   Centre, Ole Maal\o es vej 3, 2200 Copenhagen~N, Denmark}
\affiliation[b]{European Spallation Source ERIC, P.O. Box 176, 22100 Lund, Sweden}
\affiliation[c]{Glasgow University, School of Physics \& Astronomy, Glasgow G12 8QQ,
   Scotland, United Kingdom}
\affiliation[d]{Niels Bohr Institutet, Blegdamsvej 17, 2100 Copenhagen~\O , Denmark}
\affiliation[e]{Mittuniversitetet, 851 70 Sundvall, Sweden}
\affiliation[f]{CERN, Route de Meyrin, 1211 Gen\`eve, Switzerland}
\affiliation[g]{Division of Nuclear Physics, Lund University, Sweden}

\emailAdd{mortenchristensen@esss.se}

\abstract{
European Spallation Source (ESS) will deliver neutrons at high flux for use in diverse
neutron scattering techniques. The neutron source facility and the scientific instruments
will be located in Lund, and the Data Management and Software Centre (DMSC),
in Copenhagen.
  A number of detector prototypes are being developed at ESS together with its European
in-kind partners, for example: SoNDe, Multi-Grid, Multi-Blade and Gd-GEM. These are all
position sensitive detectors but use different techniques for the detection of
neutrons.
  Except for digitization of electronics readout, all neutron data is anticipated
to be processed in software. This provides maximum flexibility and adaptability
\red{and allows deep inspection of the raw data for commissioning which will reduce
the risk of starting up new detector technologies.}
But \red{it} also
requires development of high performance software processing pipelines and optimized
and scalable processing algorithms.
  This report provides a description of the ESS system architecture for the neutron
data path. Special focus is on the interface between the detectors and DMSC
which is based on UDP over Ethernet links.
  The report also describes the software architecture for detector data processing
and the tools we have developed, which have proven very useful for efficient early
experimentation, and can be run on a single laptop.
  Processing requirements for the SoNDe, Multi-Grid, Multi-Blade and Ge-GEM
  \red{detectors are presented and compared to event processing rates archived so far.}
}


\begin{document}
\maketitle
\flushbottom

\section{Introduction}
The European Spallation Source \cite{ess1, ess2} is a spallation neutron source currently being built
in Lund, Sweden.  ESS will initially support 15 different instruments for neutron scattering.
The ESS Data Management and Software
Centre (DMSC), located in Copenhagen, provides infrastructure and computational support for the acquisition,
event formation, long term storage, and data reduction and analysis of the experimental data.
At the heart of each instrument is a neutron detector and its associated readout system. Currently detectors as
well as readout systems are in the design or implementation phase and various detector prototypes
have already been produced \cite{sonde1, multigrid1, multiblade1, gdgem1}. ESS detectors will operate in event
mode\cite{gahl}, meaning that for each detected neutron a (time, pixel) tuple is calculated, providing the detection
timestamp (with a resolution of 100 ns or better) and position on the detector where the neutron hit. This allows for later filtering of individual events (vetoing) and flexible refinement of the energy determination as well as of the scattering vector.

ESS detector prototypes have been tested at various neutron facilities
and a number of temporary data acquisition systems have been in use so far.
When in operation, ESS will use a common readout system which is currently being
developed \cite{kolya1}.
We are also moving towards a common software platform for the combined activities of data acquisition
and event formation. This platform consists of core software functionality common to all detectors and a
detector specific plugin architecture.

The main performance indicators of the system are: the neutron rates, the data
transport chain from the front-end electronic readout to the event formation system, the parsing requirements
for the readout data, and the individual data processing requirements for the different detector technologies.

Good estimates of the neutron flux on the sample and the detectors have been produced by
simulations \cite{simtools, irina, kelly}, and estimates on the corresponding data rates
have been made, although the precise values will depend upon engineering design decisions that
are still to be made or further detector characterizations. Examples of these are: number of triggered
readouts per neutron, readout data encapsulation methods, hardware data processing, etc.

\red{
The architecture for the ESS data path is described in section \ref{sec:essarch}, and
section \ref{sec:readout} briefly describes ESS readout architecture
and discusses hardware and physical abstractions such as digital and logical geometry.
  Parsing of readout data and
event formation processing is the subject of section \ref{sec:detectors}, which
describes four detectors (sections \ref{sec:sonde} - \ref{sec:gdgem}), which
have been subjected to early testing at neutron sources
using a scaled-down version of the anticipated software infrastructure for ESS operations.
The software architecture is described in section \ref{sec:efu}, where the choice and
performance of UDP for detector readout is discussed in section \ref{sec:udpperf}.
}

\red{Throughput rates} for event processing are reported in section \ref{sec:performance}.

\subsection{Instrument data rates}
ESS is a spallation source, where neutrons are generated by the collision of high energy protons with a
suitable target -- tungsten in case of ESS.
The proton source is pulsed, with a 14 Hz frequency. The neutron flux generated by
this process has been simulated with MCNP from the target/moderator \cite{mcnp, mcnpb, moderator}.
The flux is reduced throughout the neutron path by neutron transport components (neutron guides,
monitors, beam ports) and instrument-specific components (collimators, choppers, sample
enclosures, etc.) and are typically calculated using the  Monte Carlo simulation tool
McStas \cite{mcstas, mcstasb}. The detector properties are determined by a combination of Geant4 simulations
\cite{geant4, geant4b, geant4c} for the initial considerations and experiments at neutron facilities
once a prototype has been built.

  Typically, rates are reported for neutrons hitting the sample and the detector, as
shown in table \ref{tab:instrument_rates}.
  As mentioned, these rates are not directly convertible into data rates received by the software. Our estimates
of the required processing power are, however, based on neutron rates for the detector surface assuming
100\% efficiency. On the input side of software processing we use the peak instantaneous rate, measured as the
highest number of neutrons received in a 1~ms time bin \cite{irina}, because we must receive all data without loss.
On the output side, we use average rates because event data is buffered up for transmission inside the event
formation system which has the practical effect of load levelling the event rate with time.

\begin{table}[tbp]
  \footnotesize
  \centering
  \caption{\label{tab:instrument_rates} Current estimates of data rates for selected ESS instruments
    at 5MW ESS source power. Global avg. rate is defined in \cite{irina}. Data Rate is the corresponding
    amount of data received for software processing.}
  \smallskip
  \begin{tabular}{llcccc}
    \Xhline{2\arrayrulewidth}
      \textbf{Instrument} &  \textbf{Detector}   & \textbf{Flux on sample}   &\textbf{Global avg. rate}  & \textbf{Data Rate} \\
                 &             &  [$\mathrm{n/s/cm^2}$ ]    &   [MHz]     &  [MB/s]   \\
    \hline
      C-SPEC             & Multi-Grid  & $10^8$      &   10    &     80  \\ 
      ESTIA              & Multi-Blade & $10^9$      &  500    &   4000  \\ 
      FREIA              & Multi-Blade & \e{5}{8}    &  100    &    800  \\ 
      LoKI\footnotemark  & BAND-GEM    & $\leq 10^9$ &   34    &    272  \\ 
      LoKI               & BCS         & $\leq 10^9$ &   37    &    298  \\ 
      NMX                & Gd-GEM      &\e{4.8}{8}   &    5    &    300  \\ 
      SKADI              & SoNDe       & $\leq 10^9$ &   37    &   1180  \\ 
      T-REX              & Multi-Grid  & $10^8$      &   10    &     80  \\ 
    \Xhline{2\arrayrulewidth}
  \end{tabular}
\end{table}
\footnotetext{The detector technology for LoKI has not been finally decided.}

\subsection{State of the art}
\red{At most existing neutron facilities, the data acquisition is highly beam line specific, and based around the DAQ PC on the beamline, as the instrument
control and DAQ PC, for example} \cite{IBR2-DAQ}. \red{It can not be seen as an integrated system at the facility level. Part of the reason for this is to be able
to continue operation of legacy systems. Additionally this is the simplest approach to DAQ.}

\red{Starting with ISIS, there have been moves to integrate the DAQ more at the facility level} \cite{ISIS-DAE2, ISIS-DAE2_II}. \red{More recently, the ILL has been investigating
moving some of the data reduction and processing into the online data acquisition and control} \cite{ILLDAQ}. \red{The SNS started with a DAQ system that used a
in-house custom code for instrument control and data acquisition} \cite{SNS-DAS}. \red{More recently, the SNS has upgraded most beam lines with a more integrated
and standardised DAQ and controls system. The new DAQ is controlled by EPICS directly} \cite{SNS-DAQ}.

\red{One particular feature of ESS instruments is that for the detector systems there is a dramatic increase in channel count. Typical DAQ systems for neutron
instruments at existing facilities have less than 1000 electronics channels. ESS instruments will in general have >1000 electronics channels, with some
instruments being in the 10 000- 100 000 electronics channels range. This means that a fully integrated DAQ system is needed. Additionally, the increase
in capability for networking, processing and data handling, means that many operations and algorithms that would only have been possible to do previously
either offline, after data storage, or in dedicated fast electronics, can now be done in realtime with sufficiently low latency in a single or small cluster
 of PCs} \cite{CMS-DAQ,ZEUS-GTT}. \red{The architecture foreseen for the ESS data acquisition, described in this paper, takes advantage of this capability.}

\subsection{Architecture for the ESS data path}
\label{sec:essarch}
\begin{figure}[tbp]
\centering
\includegraphics[width=1.0\textwidth,clip]{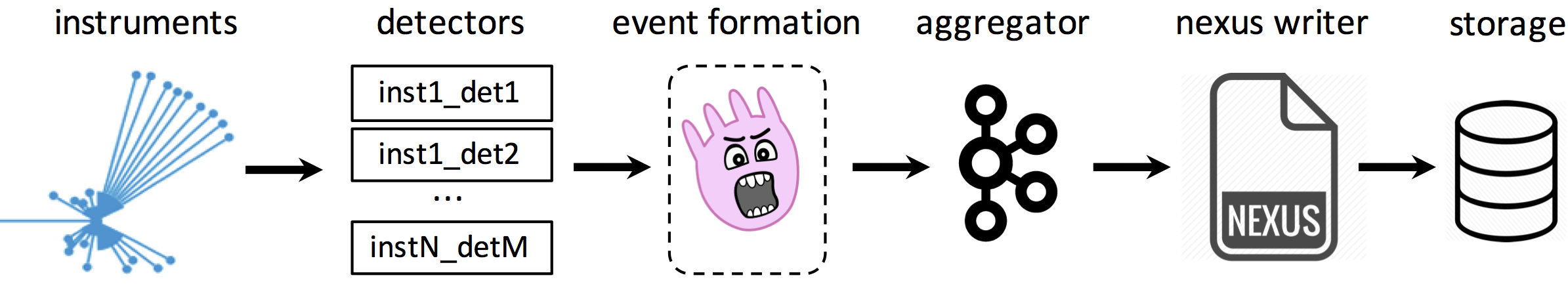}
\caption{\label{fig:ess_data_path} ESS data path.}
\end{figure}

The system architecture for the ESS neutron data path is shown in figure \ref{fig:ess_data_path}.
Every neutron scattering instrument has at least one detector: the individual
detector technologies vary \cite{kirstein}, and this is discussed in section \ref{sec:detectors}, but eventually an
electric signal is induced on an electrode. This signal is digitised by the readout electronics and
sent via UDP to the event formation system.

  The key component for the software event formation system is the Event Formation Unit (EFU), a user-space Linux
application targeted to run on Intel x86-64 processors written in C++. For each ESS instrument, several EFUs will
run in parallel to support the high data rates.
  The EFU is responsible for processing the digitised readouts and converting these into
a stream of event (time, pixel) tuples.

The event tuples are serialised and
sent to a scalable data aggregator/streamer providing a publish/subscribe
interface. A file writer application subscribes to the neutron data stream and
streams from other sources such as motor positions for collimators and sample, temperature,
pressure, magnetic/electric fields, etc. This aggregated data is then written to file in a format
suitable for long-term storage. From permanent storage
it is then possible to perform offline data reduction and analysis \cite{afonso}.
\red{The data streams are not only available to file writing, but will also be used
for live data reduction by Mantid \cite{mantid}.}

\section{Detector readout}
\label{sec:readout}
\begin{figure}[tbp]
  \centering
  \includegraphics[width=0.7\textwidth,clip]{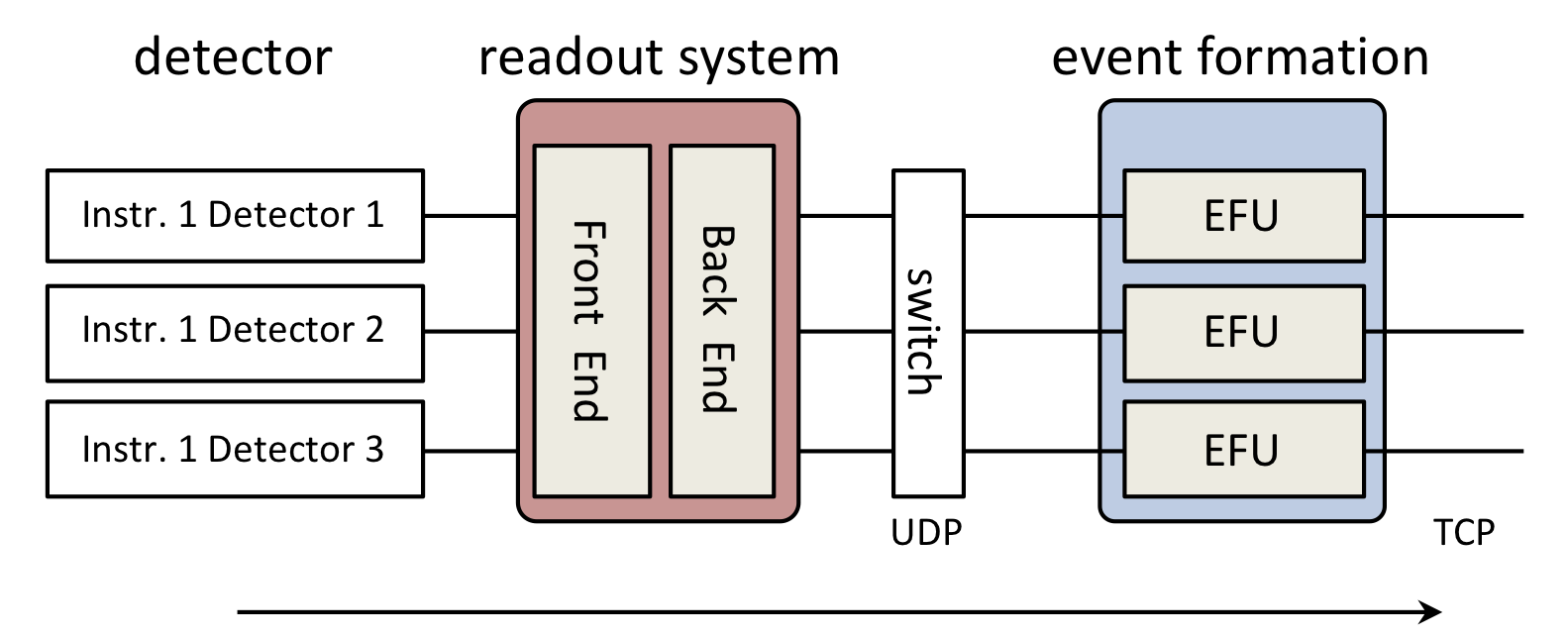}
  \caption{High level architecture of the ESS readout system.}
  \label{fig:ess_readout}
\end{figure}

The subject of this report is mainly concerned with the detector data flowing from
the readout system backend to and through the event formation system. Due to
the high neutron flux delivered by ESS, the data rates will be correspondingly high. The ESS
readout system conceptually consists of a detector specific front end and a generic backend as illustrated
in \red{figure} \ref{fig:ess_readout}.

The back end
connects to the event formation system via 100~Gb/s optical Ethernet links, which
provide more capacity than required for most instruments. However,
for the small scale detector prototypes we typically use Gigabit Ethernet.

The ESS readout system is currently under development. Until it becomes generally
available a number of different ad-hoc readout systems have been employed
for testing of prototypes. The ones relevant
for this report are: CAEN, mesytec, RD51 Scalable Readout System and ROSMAP-MP from
Integrated Detector Electronics AS
\cite{caen, mesytec, srs, ideas}.
These are either controlled by applications supplied by the manufacturer, by
custom Python scripts or GUI applications. The digitised data is transmitted as binary
data over UDP in a similar way as when the instruments are in operation.
None of these ad-hoc systems is currently set up to consume the ESS absolute timing information.

\subsubsection*{Digital geometry}
While the common readout back-end deals with the connection to the event formation,
different detector technologies have different electrical connections to the readout
front ends. Multi-Grid for example, uses a combination of wires and grids whereas Multi-Blade uses
wires and strips. Even for a specific detector technology, the different prototypes can have different sizes
and therefore different number of channels.
We need to combine the knowledge about the electrical wiring and how the digitisers
are connected in order to know anything about where on the detector a signal was induced. We call
this the digital geometry.

\begin{figure}[tbp]
  \centering
  \includegraphics[width=1.00\textwidth,clip]{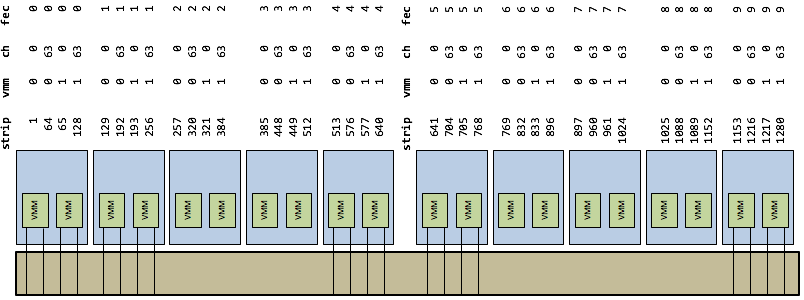}
  \caption{\label{fig:gdgem_digital_geometry} An example of a possible mapping of the digital
  geometry for x-strips for Gd-GEM. Strip 1 corresponds to asic 0, channel 0, front end card id 0 and strip
  1280 to  asic 1, channel 64, front end card id 9.}
\end{figure}

An example outlining the digital geometry for Gd-GEM is shown in
figure \ref{fig:gdgem_digital_geometry}. The x-position is a function $x(a, c, f)$ of an \textbf{asic id}
ranging from 0 to 1, a \textbf{channel} from 0 to 63 and a \textbf{front end card id} from  0 to 9.
  For each detector pipeline, a digital geometry C++ class is created
to handle this mapping. The classes are typically parametrised so they can handle multiple variants.

\subsubsection*{Logical geometry}
\begin{figure}[htbp]
  \centering
  \includegraphics[width=.4\textwidth,clip]{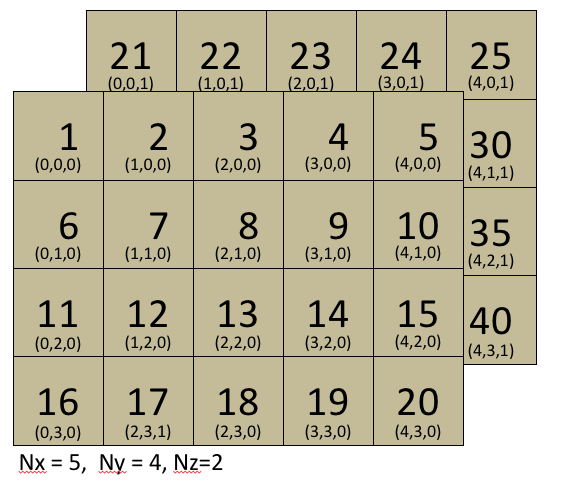}
  \caption{Logical geometry convention for 3D detectors. The example shows a single-panel 3D
    detector geometry with dimensions nx, ny, nz = (5,4,2), and pixel id's from 1 to 40.}
  \label{fig:logical_geometry}
\end{figure}

The main end result from the event formation are event tuples. An event tuple (t,p) consists of
a timestamp and a pixel\_id. Due to its physical construction, the detectors are inherently
pixellated and what we calculate is simply which pixel was hit by a neutron, i.e. this
step does not need to know anything about the physical size or absolute compositions
of the pixels. We call this the logical geometry, \red{the principle is illustrated in figure \ref{fig:logical_geometry}.}
%
We have defined a common
convention for the logical geometry for ESS instruments. The convention covers
single-panel and multi-panel, 2D and 3D detectors. For example Multi-Grid is a
single panel 3D detector (which then has voxels instead of pixels, but we do not
make a distinction) and Gd-GEM is a multi-panel 2D detector. In this scheme we
also unambiguously define the mapping between the (x,y,z) coordinates of the (logical) positions
and a unique number, called the pixel\_id.

\section{Four ESS detector technologies}
\label{sec:detectors}
Neutrons cannot be directly observed, but are observed as
the result of a conversion event where the neutron interacts
with a material with a high thermal neutron cross section for absorption. In
this process the absorber material converts the neutron into charged particles
or light, which can then be detected by conventional methods. For the
detectors in this study the conversion materials are based on
$\isotope{Li}$, $\isotope{Be}$ and $\isotope{Gd}$. The detection methods for
the individual detectors will be described below.

\subsection{SoNDe}
\label{sec:sonde}
\begin{figure}[tbp]
  \centering 
  \includegraphics[width=1.0\textwidth,clip]{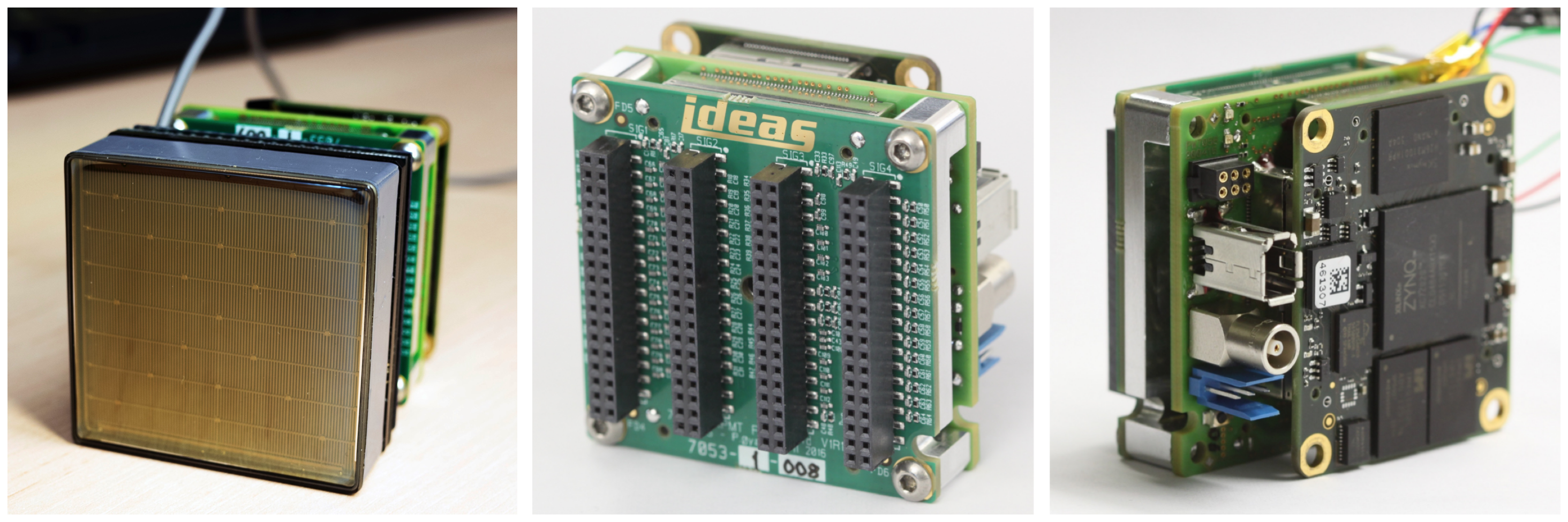}
  \caption{\label{fig:sonde_module} Single SoNDe module with PMT (left), ROSMAP module PMT interconnect side (middle),
  and digital electronics (right). Photos: European Spallation Source.}
\end{figure}

The Solid-state Neutron Detector (SoNDe) is based on a scintillating material that
converts thermal neutrons into light which is detected by a photomultiplier tube.
The detector is in an early stage of characterisation and is
currently available as a single module demonstrator \cite{sonde1}, shown in
figure \ref{fig:sonde_module}. It consists of a pixelated scintillator, a Hamamatsu H8500 series 8 x 8 MaPMT
 (Multi-anode Photomultiplier Tube) \cite{h8500pmt} and a SONDE/ROSMAP-MP counting chip-system to read out the MaPMT
signals \cite{rosmap}. The chip-system consists of four ASICs each responsible for readout of
16 pixels. The final detector will consist of 400 of such
modules, arranged in 100 groups of four modules in a 2 by 2 configuration. For a report on the
recent progress and patent information, see \cite{sonde2, sonde3}.

The ROSMAP
module transmits readout data in three different operation
modes as UDP data over Ethernet. The supported modes are Multi-Channel Pulse-Height Data,
Single-Channel Pulse-Height Data and Trigger Time Hits over threshold Data.
For early characterisation and verification it is necessary
to extract the charge information for individual channels and thus support for
the two ``expert mode'' data formats have been developed. When in operation at ESS only the
event-mode format (Trigger Time) will be relevant.

\subsubsection*{Processing requirements}
SoNDe belongs to a class of detectors requiring little data processing as the readout system
already provides event data in the form of (time, asic{\_}id and channel) values. The digital
geometry only has to account for the fact that two of the readout ASICs are rotated 180 degrees
compared with the others and the fact that they represent a view of the detector surface
from the back, which is different from the logical geometry definition we use.
  For the single module demonstrator, which consists of 8 x 8 pixels the processing steps are
\begin{itemize}
  \itemsep0em
  \item parse the binary readout data and extract (time, asic{\_}id, channel)
  \item combine asic{\_}id and channel to a pixel{\_}id
\end{itemize}

\subsection{Multi-Grid}
The Multi-Grid (MG) detector has been introduced at ILL and developed in a
collaboration between ILL, ESS and Link\"oping University. The detector is based on thin
converter films of boron-10 carbide \cite{hoglund, hoglundb} arranged in layers orthogonal to the incoming
neutrons.
  The MG detector uses a stack of grids with a number of wires running
through them.

\begin{figure}[tbp]
  \centering
  \includegraphics[width=0.8\textwidth,clip]{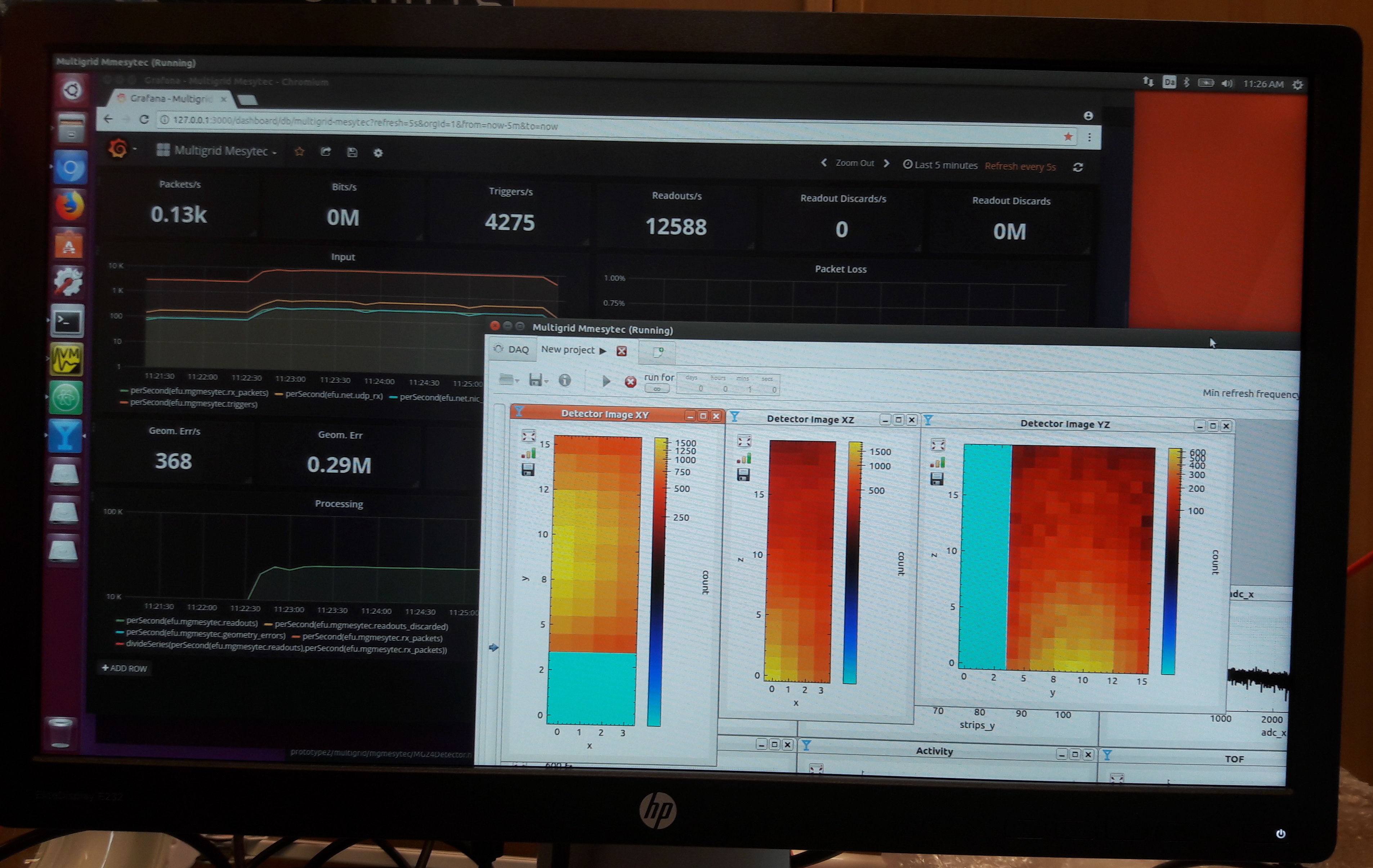}
  \caption{\label{fig:multigrid_daq} Grafana dashboard and live detector images from a
  recent test run with low neutron intensity at the Source Testing Facility at Lund University \cite{messi}.}
\end{figure}

Following the neutron conversion, a signals are induced both on grids
and wires, which are digitised and read out. The temporal and spatial
coincidence of the signals on wires
and grids is used to determine neutron positions. Signals can be induced on multiple
grids, and for double neutron events also on wires. The detector geometry
is three-dimensional, so our visualisation of the detector image consists of projections of
the neutron counts onto the xy-, xz-, and yz-planes respectively as shown in figure \ref{fig:multigrid_daq}.

\subsubsection*{Readout}
The Multi-Grid readout system used for prototyping and demonstration
detectors is based on stacked MMR readout boards supporting 128 channels,
a Mesytec VMMR-8/16 VME receiver card supporting up to 16 readout links, and a SIS3153 Ethernet
to VME interface card. It is self-triggered: when the Mesytec hardware registers a signal
above a certain trigger-threshold, it triggers a readout of all channels with signals above a
second threshold. This readout is
then transmitted as UDP packets to the EFU. The binary data format is hierarchical as it supports
multiple interface cards, each supporting multiple boards with up to 128 channels.

\subsubsection*{Processing requirements}
The Mesytec UDP protocol has been partially reverse-engineered based on captured network
traffic and the available documentation. The protocol parser must be able to support
multiple triggers in a single packet, and to discard unused or irrelevant data fields. The data fields
consist of 32 bit words each containing a command (8 bits), address/channel (12 bits) and ADC values (12 bits).
The channel readouts are given in alternating order (1, 0, 3, 2, 5, 4, ...).
All channels are assigned a single common 32-bit timestamp, in units of 16 MHz ticks, by the
electronics. Thus temporal clustering is performed in hardware, but no continuous global time
is currently available.

The EFU then parses the channel readouts, and applies software thresholds. At
this stage it discards inconsistent readouts. Channel readouts for Multi-Grid
are then mapped to either a grid or a wire id. The current algorithm for the Multi-Grid event
formation simply uses the maximum ADC values for grids and wires to determine the position. The
processing steps thus consist of

\begin{itemize}
  \itemsep0em
  \item parse the binary Mesytec readout format to extract time, channel and ADC
  \item discard inconsistent readouts
  \item map channel to either grids or wires
  \item apply suppression thresholds independently for wires and grids
  \item check for coincidence (must involve both one wire and one grid)
  \item combine wire{\_}id and grid{\_}id to pixel{\_}id
\end{itemize}

\subsection{Multi-Blade}
The Multi-Blade detector is a stack of Multi Wire Proportional Chambers operated
at atmospheric pressure, with a continuous gas flow. It consists
of a number of identical units, called cassettes. Each cassette holds a blade
(a substrate coated with $\isotope[10]{B}_4\isotope{C}$ ) and a two-dimensional readout system, which
consists of a plane of wires and a plane of strips. The cassettes
are arranged along a circle-arc centered on the sample, and are angled slightly
with respect to the neutron beam, for improved counting rate capability and spatial resolution.
The operation is based on the temporal and spatial coincidence of signals on strips and wires.
  Despite inherently being a three-dimensional detector, the visualisations of the detector images
typically display an ``unfolded'' two-dimensional pixel map. For further details of the design and
performance of this detector see \cite{multiblade2, multiblade4, multiblade5}.

\subsubsection*{Readout}
The Multi-Blade detector prototype currently has nine cassettes, each with 32 wires and 32 strips, for a
total of 576 channels. The readout is based on six CAEN V1740D digitisers, and a custom readout
application based on the API and software libraries supplied by CAEN. The digitisers each
have 64 channels, 32 for wires and 32 for strips. The wires and strips are connected to the
digitiser via front-end electronics boards. The final detector will have 32 wires and 64
strips per cassette, and up to 50 cassettes for a total of 4800 channels.

When the CAEN readout
system has detected signal above a certain (hardware) threshold it triggers an individual
readout of that channel. The readout consists of a channel number, a pulse integral (QDC),
a time-stamp, and digitiser id. For each trigger there will be one or more signals from both wires and strips.
The readout application continuously reads from the CAEN digitiser's hardware registers
using optical links and transmits the raw data over UDP to the event formation unit.

\subsubsection*{Processing requirements}
Readouts are subject to clustering analysis, where they are matched in both time and amplitude. The
maximum timespan for which channels can be said to belong
to the same cluster is a configurable parameter of the algorithm. For coincidence building
there can be up to 2 wires and
4 strips in a cluster, where the typical case is  one wire and two strips.
   Following clustering, we then calculate the pixel where the neutron was detected and
adds a timestamp. To summarise, the processing steps for Multi-Blade are:

\begin{itemize}
  \itemsep0em
  \item parse the UDP readout format to extract time, digitiser, channel and QDC values
  \item collect readouts in clusters
  \item map channel ids to either strips or wires
  \item check for coincidence (time and amplitude)
  \item combine wire{\_}id and strip{\_}id to pixel{\_}id
\end{itemize}

It is possible to improve the spatial resolution by employing CoG (center of gravity) on strip
readouts weighted by the deposited charge (QDC).

The measured amplitudes on the wires and on the strips are strongly correlated. This means that with
sufficient dynamic range double neutron events, which would cause some ambiguity, might be resolved
by requiring matching amplitudes \cite{multiblade2}.

  The processing pipeline for Multi-Blade currently differs from the other
detectors in that the code responsible for clustering and event formation runs in multiple
incarnations, namely one for each cassette. This is a case
where we explore the solution space for event processing. The approach has the advantage of supporting
individual processing for each blade rather than having to explicitly maintain information
about blade id's in the processing algorithm itself.

\subsection{Gd-GEM}
\label{sec:gdgem}
\begin{figure}[tbp]
  \centering
  \includegraphics[width=0.6\textwidth,clip]{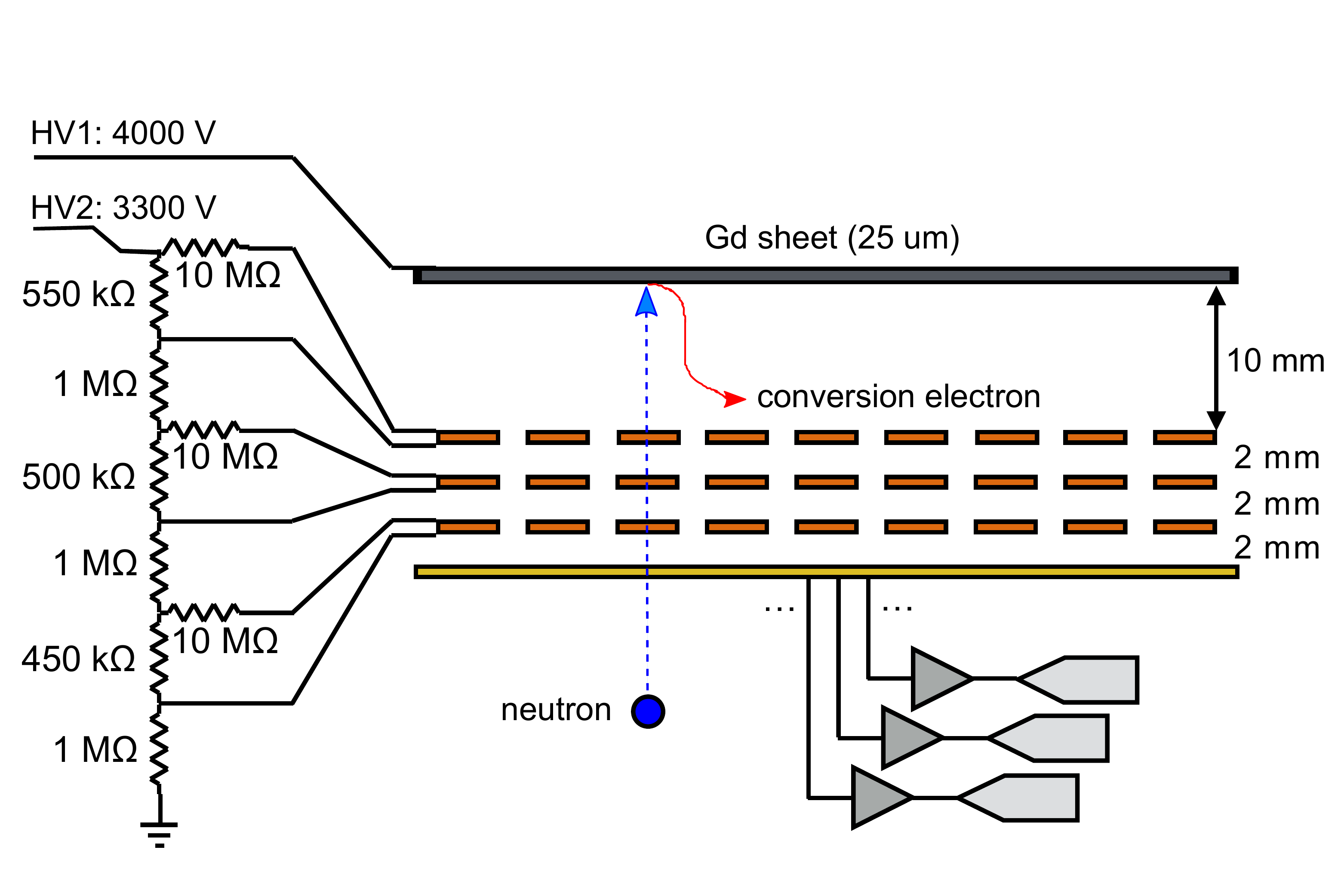}
  \caption{\label{fig:gdgem_detector} Schematic drawing of the Gd-GEM detector in backwards mode. From \cite{gdgem1}.}
 \end{figure}

The NMX macromolecular diffraction instrument will use the Gd-GEM detector technology. The neutron converter is a \SI{25}{\micro\meter} thin
foil of gadolinium, which also serves as cathode in a gas volume ($\isotope{Ar}/\isotope{CO}_2$ 70/30 at athmospheric pressure).
After traversing the readout and the GEM foils, the neutron hits the converter where it is captured as shown in figure~\ref{fig:gdgem_detector}.
After the neutron capture, gamma particles and conversion electrons are released into the gas volume.

The conversion electrons loose energy by ionizing the gas atoms, and create secondary electrons along their path.
Due to an electric field, those secondary electrons are drifted away form the cathode to an amplification stage
consisting of a stack of two or three GEM foils. Each electron generates a measurable amount of charge by an avalanche
in the GEM holes, which induces a signal on a segmented anode.

This segmentation is realised by copper strips
with a pitch of \SI{400}{\micro\meter}. The signal on the strips is read out with a timing resolution in the
order of \SI{10}{\nano\second}, such that projections of the tracks in the x-t and y-t plane can be used to
combine hits in both planes (clustering) and reconstruct the neutron impact point (micro-TPC method)~\cite{gdgem2}.

\subsubsection*{Readout}
\begin{figure}[tbp]
  \centering
  \includegraphics[width=0.7\textwidth,clip]{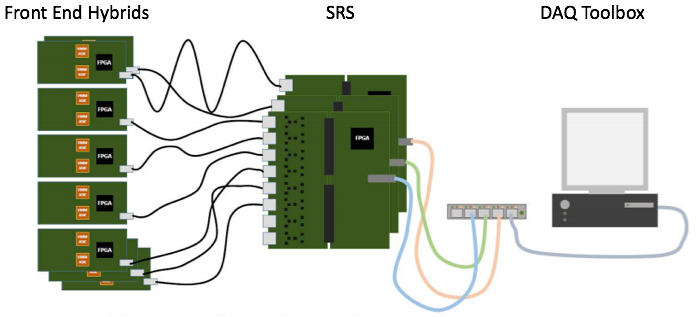}
  \caption{\label{fig:gdgem_daq_figure} Figure of the Gd-GEM readout and data acquisition system, from \cite{VMM_SRS}.}
\end{figure}

The analogue signals of the strips of the Gd-GEM detector are read out by the VMM ASIC developed by Brookhaven National Laboratory for
the New Small Wheel Phase 1 upgrade \cite{vmm3}. The VMM has been implemented in the SRS~\cite{VMM_SRS} at CERN and a schematic
drawing of the readout chain is shown in figure~\ref{fig:gdgem_daq_figure}. The so called front-end hybrids are directly mounted onto
the detector. This PCB holds two VMM ASICs, each with 64 input channels connected to the anode strips with a spark
 protection circuit. For each hit strip where the signal surpasses a configurable threshold, the VMM outputs a 38
 bit binary word, see table~\ref{tab:vmm3}.

For the prototype a Spartan-6 FPGA on the hybrid controls the ASICs and bundles the data, that are transmitted via HDMI cables to the
core of SRS, the Front-End Concentrator (FEC) card. Up to eight hybrids can currently be connected to one FEC and the
data are encapsulated into UDP packages of a 1~Gb/s Ethernet connection to the readout computer \cite{gdgem_electronics}.

The readout of the Gd-GEM detector is partitioned into 4 sectors. Each of these 4 sectors has 640 strips read out by
5 hybrids in x and y direction, resulting in a total of 5120 strips and 40 hybrids. If a signal is recorded on a
detector strip, the VMM on the hybird generates hit data for the corresponding channel. Before sending
out the data via UDP, the FEC adds the VMM ID and the FEC ID to the hit data.

\begin{table}
  \centering
  \caption{Data fields from VMM3.}\label{tab:vmm3}
  \smallskip
  \begin{tabular}{|l|r|l|r|l|r|}
    \hline
      \textbf{field} & \textbf{size (bits)} & \textbf{field} & \textbf{size (bits)} & \textbf{field} & \textbf{size (bits)} \\
    \hline
      flag & 1 & threshold & 1 &  channel & 6 \\
    \hline
      amplitude & 10 &  time & 8 &  BCID & 12 \\
    \hline
  \end{tabular}
\end{table}

With the information tuple (channel,  VMM ID, FEC ID), the geometrical position of each
hit can be reconstructed. A configuration file that reflects this digital geometry of the
detector is loaded during the start up phase of the DAQ.
The configuration can be modified for reordering, exchange or extension of physical readout components.

\subsubsection*{Processing requirements}
The Gd-GEM detector requires the most complicated processing requirements in
terms of the physical processes, the data acquisition and processing power.
  The steps required are
\begin{itemize}
  \itemsep0em
  \item parse the binary data from SRS readout and extract (time, channel, adc)-tuples
  \item queue up (time, channel, adc)-tuples until enough data for attempting clustering analysis
  \item perform clustering analysis - determine if coincidence occurred
  \item calculate neutron entry position for x and y
  \item convert positions to pixel{\_}id
\end{itemize}

Some of the software related challenges for Gd-GEM are: Scaling up to a full rate, detector size and
for discriminating invalid tracks. A neutron event generates a track with extensions in both time and
space so it is not possible to just partition the detector in regions for independent parallel
processing.
Several processing options for distinguishing which tracks from which position can be
extracted have been described in \cite{brightness1}. In addition, due to the required buffering of
data, memory usage and cache performance may well be a concern.


%
%
\section{Event Formation Unit architecture}
\label{sec:efu}

\begin{figure}[tbp]
  \centering
  \includegraphics[width=.95\textwidth,clip]{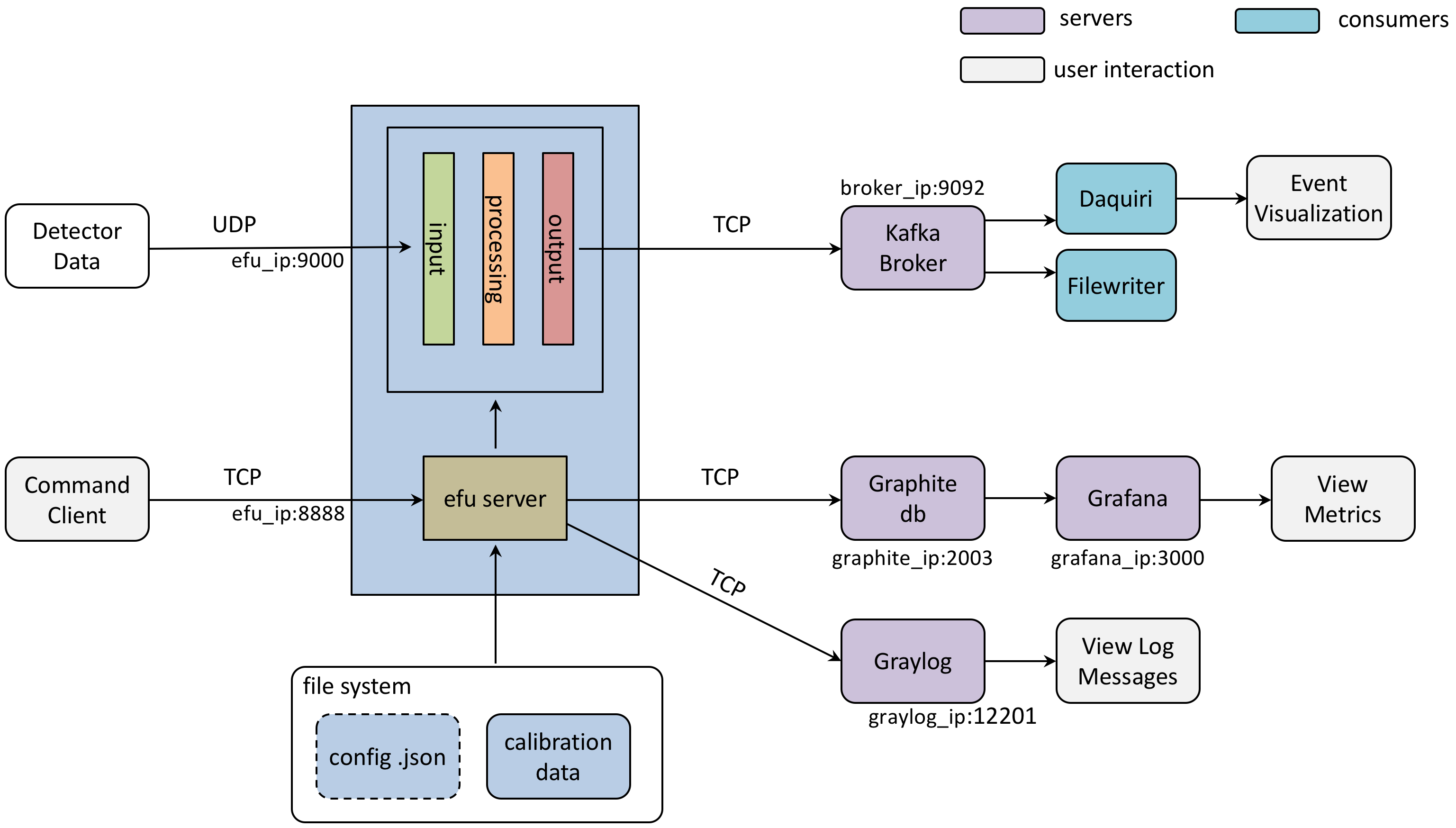}
  \caption{\label{fig:efu_architecture} Event Formation Unit (EFU) architecture.}
\end{figure}

The EFU architecture, illustrated in figure \ref{fig:efu_architecture}, consists of a main application
with common functionality for all detectors and detector-specific processing pipelines.
  The software is written in C++, and is built using gcc and clang compilers for Ubuntu, macOS and CentOS.
CentOS is currently the target Linux distribution for ESS operations, whereas the other operating systems
are used during development and implementation.

 The main application handles low CPU-intensity tasks such as launch-time configuration via command-line
options, run-time configuration using a TCP-based command API, application state logging and periodic
reporting of run time statistics and counters.

  Detector pipelines are responsible for handling realtime readout data,
and must conform to a common software interface definition. The pipelines are implemented
as shared libraries that are loaded and launched by the main application as
POSIX threads with support for thread affinity, which fixes a thread onto a specific processor core.
  The plugin must specify at least one processing thread but apart from this
no further restrictions are imposed. We have experimented with different
configurations for different detectors, but currently the number
of threads in a detector pipeline ranges from one to three.

When more than one thread is
in use, the data is shared between the producer and the consumer thread by a circular data buffer
(FIFO) which preserves the order of the arriving data. The FIFO is based on pre-allocated memory to
avoid unnecessary data copying and C++ \textbf{std::atomic}
primitives for resource locking. For performance benchmarking
we use the \textbf{rdtsc()} instruction call, which
gives a high resolution timestamp counter with low latency.

The data processing part of the detector pipelines generally consists of a tight loop with a BSD socket
\textbf{recvfrom()} system call, a \textbf{parse()} function, and a \textbf{produce()} step. These
processing steps can be done in a single or multiple threads, depending on specific requirements.

\subsection{UDP Data Input}
\label{sec:udpperf}
At ESS we have chosen User Datagram Protocol (UDP) for the transmission of readout data
over Ethernet. Other contenders were Transmission Control Protocol (TCP), 'raw' Ethernet frames
and (briefly) InfiniBand.
UDP is a simple protocol for connectionless data transmission \cite{udp1} without a guaranteed delivery.
The alternative, TCP, guarantees the
ordered delivery of data and automatically adjusts transmission to match the capability of the transmission
link \cite{tcp1}.

Despite its inherent unreliability, UDP is widely used. For example in the RD51 Scalable Readout System \cite{srs}, or the
CMS trigger readout \cite{cms1}, both using 1 Gb/s Ethernet.
The ESS readout system described in section \ref{sec:readout} is based on FPGAs with support
for transmission of Ethernet packets over 100G. Implementing TCP on these are not
an option, so UDP was eventually decided.

The unreliability of UDP is widely overestimated. It is true that a UDP packet can
potentially be dropped in any part of the
communications chain: The sender, the receiver, intermediate systems such as routers, firewalls,
switches, load balancers, etc. This makes it difficult in the general case to rely on UDP
for high speed communications. However high reliability solutions can be engineered for simple network
topologies such as the one anticipated for the ESS readout shown in figure \ref{fig:ess_readout}.
This is a simple network topology, where both the FPGA and the switch can deliver packets at wire
rate. Thus no packet loss will occur except at the receiver.

At the receiver, two types of data loss are the main causes of performance degradation: buffer
exhaustion and packet processing overhead. These are not independent as increased time spent
in processing data will increase the likelihood that the receive buffers fill up.
Nevertheless the effects can be reduced as we describe in the following two sections.

\subsubsection{Receive buffers}
The main parameters for controlling socket buffers are \texttt{rmem\_max}
and \texttt{wmem\_max}. The former is the size of the UDP socket receive buffer, whereas the latter is the
size of the UDP socket transmit buffer. To change these values from a BSD socket application
use \texttt{setsockopt()}, for example

\begin{verbatim}
int buffer = 4000000;
setsockopt(s, SOL_SOCKET, SO_SNDBUF, buffer, sizeof(buffer));
setsockopt(s, SOL_SOCKET, SO_RCVBUF, buffer, sizeof(buffer));
\end{verbatim}

In addition there is an internal queue for packet reception whose size (in packets) is named
\texttt{netdev\_max\_backlog}, and a network interface parameter, \texttt{txqueuelen} which were
also adjusted for our benchmark testing.

The default value of these parameters on Linux are not optimised for high speed data links such as
10 Gb/s Ethernet, so for the measurements presented here the following parameters were used:

\begin{verbatim}
net.core.rmem_max=12582912
net.core.wmem_max=12582912
net.core.netdev_max_backlog=5000
txqueuelen 10000
\end{verbatim}

These values were generally adopted from previous studies \cite{bencivenni} and guides \cite{nwtune}.

\subsubsection{Packet sizes}
Packets arriving at a data acquisition system are subject to a nearly constant per-packet processing
overhead. This is due to interrupt handling, context switching, checksum validations and header
processing. There is an intimate relation between Ethernet link speed, packet size, packet rates and
header overhead as shown in table \ref{tab:i}. For 10G Ethernet at up to 15~M packets per second,
this processing alone can consume most of the available
CPU resources.

In order to achieve maximum performance, data from the electronics readout should
be bundled into jumbo frames if at all possible. Using the maximum Ethernet packet size of 9018~bytes
reduces the per-packet overhead by up to a factor of 100. This does, however, come at the cost of larger
latency. For example the transmission time of 64 bytes is 67~ns, whereas for 9018
it is 7230~ns. For applications sensitive to latency a tradeoff must be made between low packet
rates and low latency. For ESS latency is not an issue as all readout is time-stamped by
hardware before transmission.

By bundling readout data into Ethernet Jumbo frames of size 9018 bytes, rather than using
small Ethernet frames of size 64 bytes, the packet rate is reduced by a factor of 100 as shown in
table \ref{tab:i}. This is directly measurable as a reduced time spent in system calls and reduced number
of context switching between user space and kernel.

\begin{table}[tbp]
  \centering
  \caption{\label{tab:i} Packet rates as function of packet sizes for 10 Gb/s Ethernet}
  \smallskip
  \begin{tabular}{|l|cccccccc|}
    \hline
      User data size [B] & 1  &       18 &   82 &  210 &  466 &   978  & 1472 & 8972 \\
    \hline
      Packet size [B]    & 64 &       64 &  128 &  256 &  512 &  1024  & 1518 & 9018 \\
    \hline
      Overhead [\%]      & 98.8 &   78.6 & 44.6 & 23.9 & 12.4 &   5.5  &  4.3 &  0.7 \\
    \hline
      Frame rate [Mpps]  & 14.88 & 14.88 & 8.45 & 4.53 & 2.35 &  1.20  & 0.81 & 0.14 \\
    \hline
  \end{tabular}
\end{table}

We configured the systems with an MTU of 9000 bytes allowing user payloads up to 8972~bytes when
taking into account that IP and UDP headers are also transmitted. Given the efficiency gained by
using large packets, there was no need to consider InfiniBand or raw Ethernet
to reduce the size of protocol headers, thus confirming the choice of UDP.

\subsubsection{Performance}
For an early validation of the use of UDP we ran a series of performance tests using
an experimental configuration described in \red{appendix \ref{sec::testbed}.
The testbed} consists
of two hosts, one acting as a UDP data generator and the other as a UDP receiver. As mentioned
the readout transmitter at ESS will be based on FPGA and not a Linux server. However this
does not affect the results we measured for the receiver.

\begin{figure}[tbp]
  \centering
  \includegraphics[width=0.65\textwidth,clip]{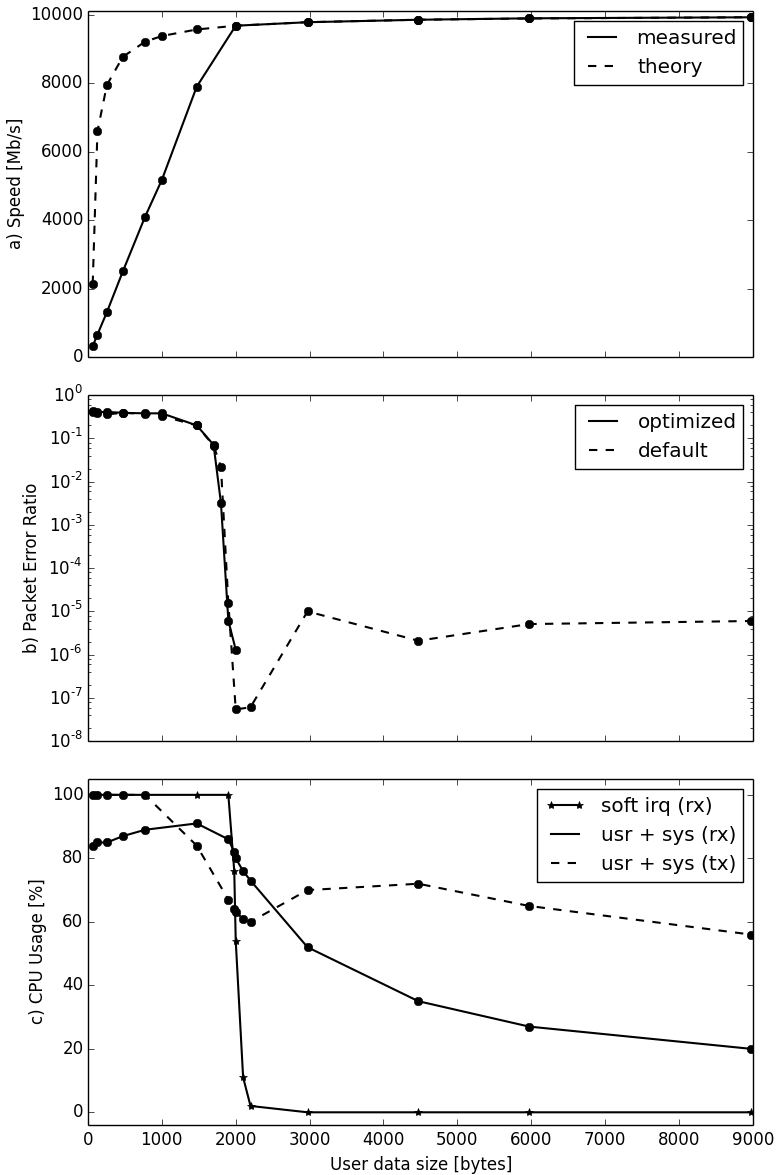}
  \caption{\label{fig:combined} Performance measurements. a) User data speed. b) Packet Error Ratio. c) CPU Load.
  Note that for the optimized values PER is zero for user data larger than or equal to 2200 bytes (solid line).}
\end{figure}

The measured performance, shown in figure \ref{fig:combined}, covers user data speed, packet error
ratios and CPU load. They are time averaged over 10 second intervals while transmitting 400~GB of
data at a time. There is a clear variation with packet size for all parameters with the
best results obtained with packet sizes larger than 2200 bytes. The best \red{results} in terms
of CPU bandwidth is obtained with a packet size of 9000 bytes.

The tests made use of packet sequence numbers allowing the determination of packet error rates (PER).
Sequence numbers are not supported in the current prototype readout systems. Thus packet
loss and PER numbers are not available for these. Sequence numbers will, however, be part of the
ESS readout system.

The main conclusion is that with only a single cpu core used as the receiver it is possible to support
10Gbit/s with zero packet loss using UDP. The actual performance when real data is applied will
naturally change, and will need to be followed closely.

\subsection{FlatBuffers/Kafka}
The output of the EFU is a stream of events. We have chosen
Apache Kafka \cite{kafka} as the central technology for transmission, and Google FlatBuffers
\cite{flatbuffers} for serialisation.
Apache Kafka is an open source software project for distributed data streaming. Multiple Kafka brokers form a
scalable cluster,
which supports a publish-subscribe message queue pattern with configurable data persistence.

In Kafka, producers send data to a topic in a cluster. A consumer subscribes to a topic to receive messages,
either from the instant the subscription starts or from a previous offset, provided the requested data is
still available in storage on the cluster, given a retention policy. Consumers may also be grouped to
distribute the processing load among different processes.

Both producers and consumers can be developed using
open source Kafka client libraries. The EFU uses \textbf{librdkafka} \cite{librdkafka}, which offers a C/C++ API.
While Kafka offers a scalable and reliable transmission of arbitrary data, FlatBuffers provides a schema-based
event serialisation method, and a mechanism for forward and backward compatibility of the schemas.
Figure \ref{lst:schema} shows the currently used schema for events.

\begin{figure}
  \lstset{basicstyle=\small\ttfamily, xleftmargin=.15\textwidth, xrightmargin=.15\textwidth}
  \begin{lstlisting}[frame=single]
include "is84_isis_events.fbs";

file_identifier "ev42";

union FacilityData { ISISData }

table EventMessage {
    source_name : string;
    message_id : ulong;
    pulse_time : ulong;
    time_of_flight : [uint];
    detector_id : [uint];
    facility_specific_data : FacilityData;
}
root_type EventMessage;
  \end{lstlisting}
  \caption{The ESS FlatBuffers event schema. The main fields of relevance are the arrays
    \textbf{time\_of\_flight} and \textbf{detector\_id} (from \url{https://github.com/ess-dmsc/streaming-data-types}).}
  \label{lst:schema}
\end{figure}

\subsection{Live detector data visualisation }
After writing the first prototype for event formation, it became clear that it would
be beneficial to also use the EFU as a DAQ system for early detector experiments and
commissioning. One of the easiest ways to validate the event processing
is to visualise the detector image and other relevant data, such as channel intensities
and ADC distributions.

\begin{figure}[tbp]
  \centering
  \includegraphics[width=.95\textwidth,clip]{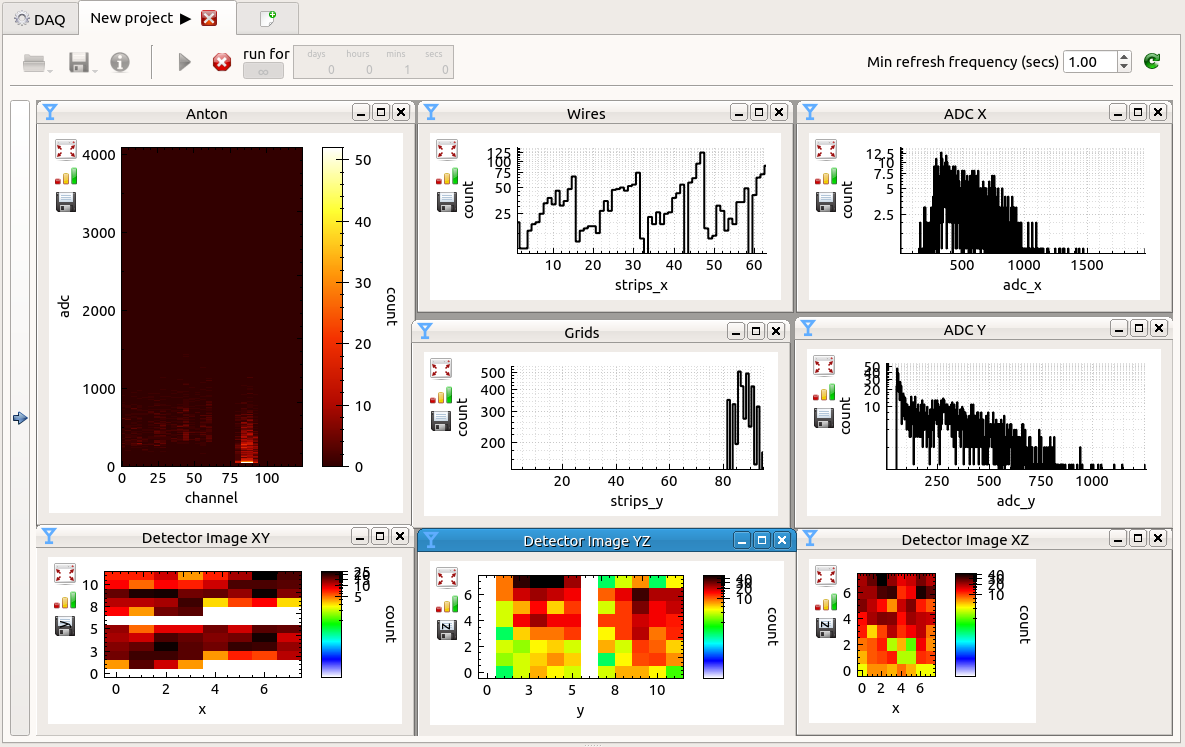}
  \caption{\label{fig:daquiri} The Daquiri commissioning tool.}
\end{figure}

For this reason the EFU also publishes such information via Kafka, and
an application named Daquiri was written for visualising these data. Daquiri
subscribes to Kafka topics, collects statistics, and provides plotting functionality, and
is planned to be an integral part of the software bundle developed for ESS operations.
The Daquiri GUI is based on Qt \cite{qt1} and is highly configurable in terms of available plotting formats,
the dashboard configuration, labels, axes, colour schemes, etc.
\red{A typical screenshot is depicted in figure \ref{fig:daquiri}.}
Daquiri is open source software \cite{daquiri1}.

\subsection{Runtime stats and counters}
The availability of relevant application and data metrics is essential for both early prototyping
and easy monitoring while in operation, for example incoming packet rates, parsing
errors, calculated events, discarded readouts, etc. The detector API provides a mechanism
for the detector plugin to register a number of named 64-bit counters. These are then periodically
queried by the main application and reported to a time-series server.

We have
chosen Graphite as the time-series server \cite{graphite1} technology and use Grafana \cite{grafana1}
for presentation. Graphite has a simple API for submission of data, which consists of
a hierarchical name such as \emph{efu.net.udp\_rx}, a counter value and a UNIX timestamp.
\begin{figure}[tbp]
  \centering
  \includegraphics[width=1.0\textwidth,clip]{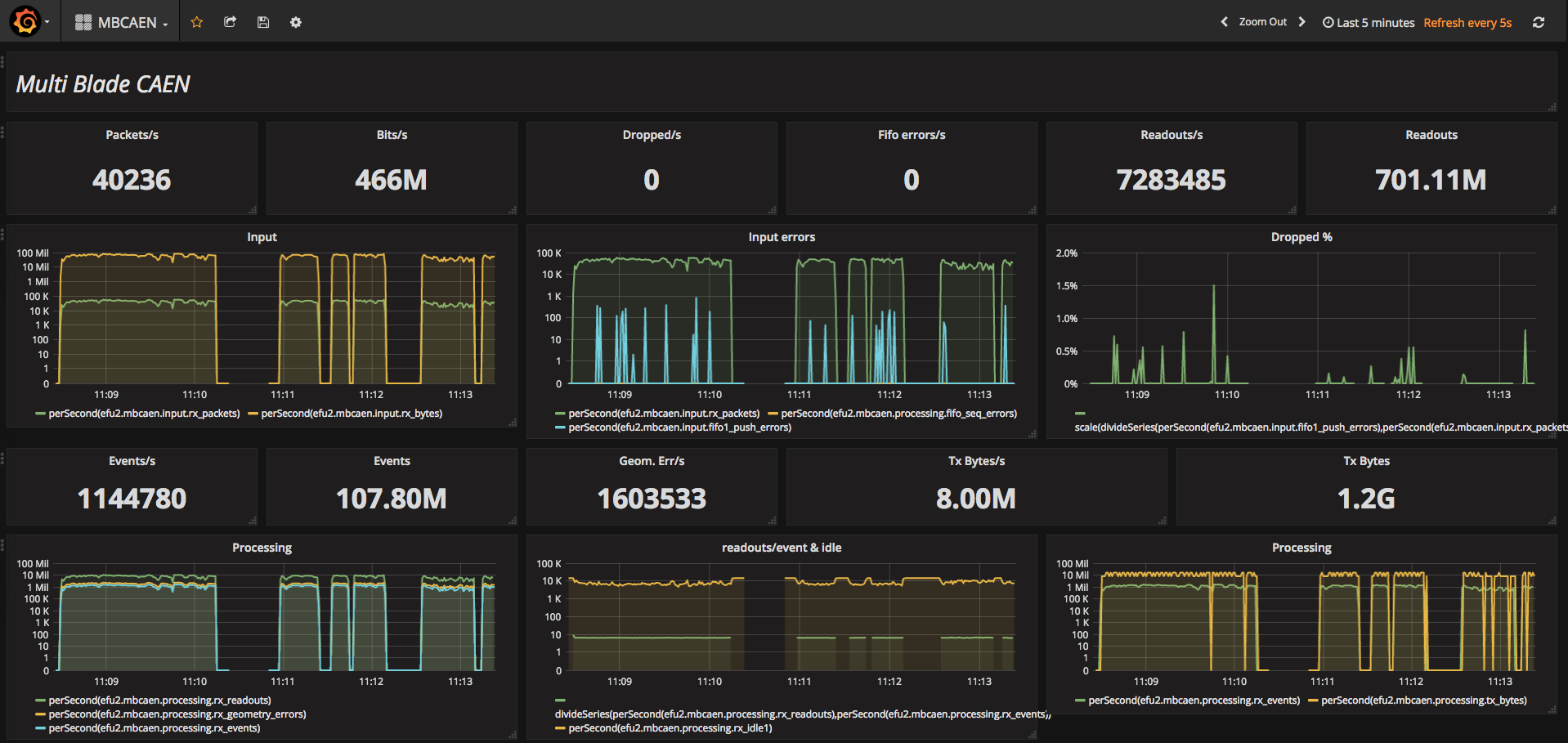}
  \caption{\label{fig:grafana_multiblade} Grafana dashboard used \red{for monitoring packet
  and event related counters for} an implementation of Multi-Blade data processing.}
\end{figure}
The combination Grafana/Graphite has proven to be very useful, not only for monitoring the event processing
software.
 Scripts have been written to check Linux kernel and network card counters, as well as
disk and CPU usage, all of which are relevant when running at high data rates while simultaneously writing
raw data to disk. We typically
publish monotonically increasing counter values, and then use Grafana to transform these into
rates. We plan to offer Grafana/Graphite for the software we are developing for ESS operations.
\red{Figure \ref{fig:grafana_multiblade} shows how this currently looks like.}

\subsection{Trace and logging}
For application logging we have chosen Graylog \cite{graylog1}. In the EFU, Graylog is used for low rate
log messages. We use the \textbf{syslog} \cite{syslog} conventions for logging levels and severities.
Graylog is not currently in wide use in the infrastructure, but will be essential for monitoring the ESS data
processing chain once in operations, when multiple EFUs are deployed.

During development we use a simple but effective
trace system consisting of groups and masks. These currently print directly to the console, which
is extremely detrimental to performance when operating at thousands of packets and millions
of readouts per second. Therefore, we made the trace macros configurable at compile time,
so that no overhead occurs when they are not needed. Both the log and trace system accept
log messages in a \textbf{printf()}-compatible format using variable arguments.

\subsection{Software development infrastructure}
All the software components that are part of the data aggregation and streaming pipeline
are being developed collaboratively by the partners as open source projects released
under a BSD license. Git \cite{git} is used for version control and all software is available for public
scrutiny on GitHub \cite{github}. We use Conan \cite{conan} as our C++ package manager and CMake
\cite{cmake} for multi-platform Makefile generation. The projects are built with gcc and clang
compilers \cite{gcc, clang}.

  A Jenkins \cite{jenkins} build server automatically triggers builds and runs commit stage tests each time
new code is pushed to a repository: Every commit on every branch for every project triggers
a Jenkins build and test cycle on multiple operating systems providing rapid feedback on breaking
changes.
  The tests that are run vary according to the application but for C++ code in general include unit
tests with Google Test \cite{googletest}, static analysis with Cppcheck \cite{cppcheck}, and test
coverage reports with gcovr \cite{gcovr}.  We also check for code format compliance with clang-format
\cite{clangformat}, and memory management problems with Valgrind \cite{valgrind}.

  Individual methods and algorithms can
be benchmarked for performance with Google Benchmark \cite{googlebenchmark}.
  The executables generated from every successful build cycle are saved as artefacts and can thus be used for
quick deployment or integration testing. Configuration of the machines in the build and test environment
is done using Ansible \cite{ansible}, with the scripts kept under version control.

\section{\red{Event processing rates}}
\label{sec:performance}

The key metric we use for the evaluation of performance is the number of events the detector
pipeline can process per second. To benchmark this, we use detector data recorded as
Ethernet/UDP packets in an number of measurement campaigns.
This data is then sent to the event formation system as fast as
possible and the achieved rates are retrieved via Grafana.

  The setup uses three servers: a macOS laptop acting as a data generator/detector readout, a Ubuntu
workstation hosting the EFU and Kafka, and another Ubuntu workstation which hosts Graphite and Grafana metrics.
  The hardware specifications are listed in table \ref{tab:perf_setup}. The tests were made on the latest event
formation software \cite{efugit}. For Gd-GEM we have implemented a performance test
based on Google Benchmark, which directly targets the event processing algorithm,
and is likely to present an upper bound for the \red{event rates} in a single processing thread
as there is no other overhead involved.

\begin{table}[tbp]
  \footnotesize
  \centering
  \caption{\label{tab:perf_setup} Machine configurations for the performance test setup.}
  \smallskip
  \begin{tabular}{llll}
    \Xhline{2\arrayrulewidth}
      \textbf{Machine}     & \textbf{OS} & \textbf{CPU} & \textbf{RAM} \\
    \hline
          detector         &  macOS 10.13.3     &  Intel Core i7 @ 2.2 GHz          &  16 GB, DDR3, 1600 MHz  \\
          efu              &  Ubuntu 16.04      &  Intel Xeon E5-2620 v3 @ 2.40GHz  &  64 GB, DDR4, 2133 MHz  \\
          metrics          &  Ubuntu 16.04      &  Intel Xeon E5-2620 v3 @ 2.40GHz  &  64 GB, DDR4, 2133 MHz  \\
          benchmarks       &  Ubuntu 16.04      &  Intel Core i7-6700    @ 3.40GHz  &  32 GB, DDR4, 2133 MHz  \\
    \Xhline{2\arrayrulewidth}
  \end{tabular}
\end{table}

Table \ref{tab:performance} summarises the results of the performance measurements. It shows that a
pipeline can support the reception and processing of around 85.000 UDP packets per second and
several millions readouts per second using one or two CPU cores. The reported event rates
reflect the amount of computational work that has to be performed on the data: Gd-GEM has the most
complex algorithm, Multi-Grid and Multi-Blade have medium complexity, and SoNDe requires the least
processing.

\begin{table}[tbp]
  \footnotesize
  \centering
  \caption{\label{tab:performance} Measured performance for detector pipelines. }
  \smallskip
  \begin{tabular}{llccccc}
    \Xhline{2\arrayrulewidth}
      \textbf{Detector} &  \textbf{Machine}     & \textbf{Packet Rate} & \textbf{Trigger Rate} & \textbf{Event Rate} & \textbf{Cores} &\red{\textbf{Packet size}} \\
                        &                       &      [ pkt/s ]       &     [readouts/s]      &   [events/s]        &       & \red{[bytes]} \\
    \hline
          Gd-GEM        &  benchmarks           &  n/a                  &      18.6 M          &   500 k -- 1 M      & 1     & \red{n/a}     \\
          Multi-Grid    &  efu                  &  86.000               &      3.0 M           &   2.46 M            & 1     & \red{1137}    \\
          Multi-Blade   &  efu                  &  82.000               &      5.6 M           &   2.31 M            & 2     & \red{1494}    \\
          SoNDe         &  efu                  &  94.000               &      23.5 M          &  23.5  M            & 2     & \red{1307}    \\
    \Xhline{2\arrayrulewidth}
  \end{tabular}
\end{table}
  The large uncertainty for Gd-GEM comes from the fact that for this detector technology neutron
events gives rise to a range of readouts of up to 20 strip hits for both x- and y-
strips. \red{None of the readout systems supported Jumbo frames at the time of these experiments,
so the additional benefit of using large packets is not reflected in the reported rates.}

Taking into account that a \red{mid-range} server can have two CPU sockets, each having 8
cores/16 hyper threads we can naively scale these numbers to the very high event rates required at ESS
by parallelisation. For example, by employing a small number (5 - 10) of servers, each dedicated to
processing data from a fraction of the detector surface, we expect to scale the rates by more than an
order of magnitude.

\section{Conclusion}
The previous sections have given an overview of the ESS software architecture
for event processing in general and as implemented in four detector designs specifically.
We have discussed the technology choices made and the toolchain used
for software development. Finally we presented recent \red{event processing rates} for four
detectors which will be used in ESS instruments.

  We have shown an architecture that can be scaled up to deal with the
high neutron rate ESS will deliver. Without having spent much time on optimisation
of the code so far we have achieved high event processing rates of the order of
1 to 25 M events per second, and have shown how this can be scaled to much higher
\red{event rates} using commodity hardware. The detectors \red{discussed in this paper represent
the range of the expected processing requirements foreseen at ESS, from simple to quite complex.}
The toolchain
does not wait to become operational after 2021 where ESS is expecting see first beam on target,
but is in actual use for data acquisition as the detector development continues.

Most detectors are constructed by the tiling of identical and independent units. Scaling the
processing up for these are easy as we can employ multiple event formation units running
in parallel.

Not all scalability problems have been solved yet, however. Future work will focus
on scaling the Gd-GEM processing as it is markedly more complicated than the other detectors.
For example a simple partitioning of the detector surface may not \red{be feasible},
because the charge tracks from a single neutron conversion
can easily cross partition borders. Collaboration on this
topic has already started. Work will also
be done on deploying multiple processing pipelines on a multi-core CPU, where typically
resource sharing problems, such as memory and network bottlenecks, will become more pronounced
than observed so far.

\acknowledgments
This work is funded by the EU Horizon 2020 framework, BrightnESS project 676548.
Ramsey Al Jebali would like to acknowledge partial support from the EU Horizon 2020 framework,
SoNDe project 654124.
  The authors would like to acknowledge the provision of beam time from R2D2 at IFE(NO), CT1 at ILL(FR),
CRISP at ISIS(UK) and the Source Testing Facility at Lund University(SE).

\vspace{5mm}
\noindent
We would also like to thank Matthew Jones, software consultant at ISIS, for his contributions on Kafka
and Google FlatBuffers.
  Finally we would like to acknowledge Kalliopi Kanaki and Irina Stefanescu, Detector Scientists at
ESS, for their comments and suggestions for improvements.



%
\appendix
\section{UDP Performance Testbed}\label{sec::testbed}
\begin{figure}[tbp]
\centering
\includegraphics[width=.6\textwidth,clip]{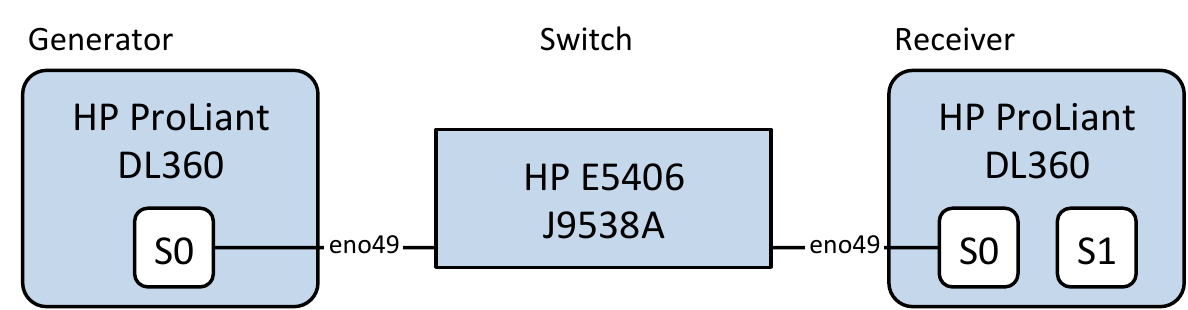}
\caption{\label{fig:expmt} Experimental setup.}
\end{figure}

For the UDP performance testing we used a two server setup shown in figure \ref{fig:expmt}.
The hosts are HPE ProLiant DL360 Gen9 servers connected to a 10 Gb/s Ethernet switch using
short (2 m) single mode fibre cables. The switch is a HP E5406 switch equipped with a J9538A 8-port SFP+ module. The
server specifications are shown in table \ref{tab:specs}. Except for processor internals the servers are
equipped with identical hardware.

The data generator is a small C++ program using BSD socket, specifically the \texttt{sendto()}
system call for transmission of UDP data. The data receiver is based on the EFU whose
architecture is described
The system, named the Event Formation Unit (EFU),
supports loadable processing pipelines. A special UDP 'instrument' pipeline was created for the
purpose of these tests. Both the generator and receiver uses \texttt{setsockopt()} to adjust
transmit and receive buffer sizes.
Sequence numbers are embedded in the user payload by the transmitter allowing the receiver to detect
packet loss and hence to calculate packet error ratios.
Both the transmitting and receiving applications were locked to a specific processor core
using the \texttt{taskset} command and \texttt{pthread\_setaffinity\_np()} function.
The measured user payload data-rates were calculated using a combination of fast timestamp
counters and microsecond counters from the C++ chrono class.
Care was taken not to run other programs that might adversely
affect the experiments. CPU usages were calculated from the
\texttt{/proc/stat} pseudofile as also used in \cite{bencivenni}.

\begin{table}[tbp]
  \small
  \centering
  \caption{\label{tab:specs} Hardware components for the testbed}
  \smallskip
  \begin{tabular}{|l|l|}
    \hline
      Motherboard                & HPE ProLiant DL360 Gen9 \\ \hline
      Processor type (receiver)  & Two 10-core Intel Xeon E5-2650v3 CPU @ 2.30GHz \\ \hline
      Processor type (generator) & One 6-core Intel Xeon E5-2620v3 CPU @ 2.40GHz  \\ \hline
      RAM                        & 64 GB (DDR4) - 4 x 16 GB DIMM - 2133MHz                      \\ \hline
      NIC                        & dual port Broadcom NetXtreme II BCM57810 10 Gigabit Ethernet \\ \hline
      Hard Disk                  & Internal SSD drive (120GB) for local installation of CentOS 7.1.1503 \\ \hline
      Linux kernel               & 3.10.0-229.7.2.el7.x86\_64 \\ \hline
  \end{tabular}
\end{table}

\section{Source code}\label{sec:code}
The software for this project is released under a BSD license and is freely available on
GitHub \cite{efugit}. To build the exact versions of the programs
used for the UDP performance experiments,
complete the steps below. To build and start the producer:
\begin{verbatim}
> git clone https://github.com/ess-dmsc/event-formation-unit
> git checkout 547b3e9
> cd event-formation-unit/udp
> make
> taskset -c coreid ./udptx -i ipaddress
\end{verbatim}
to build and start the receiver:
\begin{verbatim}
> git clone https://github.com/ess-dmsc/event-formation-unit
> git checkout 547b3e9
> mkdir build
> cd build
> cmake ..
> make
> ./efu2 -d udp -c coreid
\end{verbatim}

\end{document}